\documentstyle[12pt]{article}

\newcounter{eq}
\newcounter{sc}

\def\overleftrightarrow#1{\vbox{\ialign{##\crcr
 $\leftrightarrow$\crcr\noalign{\kern-1pt\nointerlineskip}
 $\hfil\displaystyle{#1}\hfil$\crcr}}}

\newlength{\minitwocolumn}
\begin{document}

\begin{flushright}
DPUR/TH/44\\
March, 2015\\
\end{flushright}
\vspace{20pt}

\pagestyle{empty}
\baselineskip15pt

\begin{center}
{\large\bf Hawking Radiation of a Charged Black Hole in Quantum Gravity
\vskip 1mm }

\vspace{20mm}
Ichiro Oda \footnote{E-mail address:\ ioda@phys.u-ryukyu.ac.jp
}

\vspace{5mm}
           Department of Physics, Faculty of Science, University of the 
           Ryukyus,\\
           Nishihara, Okinawa 903-0213, Japan.\\

\end{center}


\vspace{5mm}
\begin{abstract}
We study black hole radiation of a Reissner-Nordstrom black hole with an electric charge 
in the framework of quantum gravity. Based on a canonical quantization for a spherically 
symmetric geometry, under physically plausible assumptions, we solve the Wheeler-De Witt equation
in the regions not only between the outer apparent horizon and the spatial infinity but also
between the spacetime singularity and the inner apparent horizon, and then show that 
the mass loss rate of an evaporating black hole due to thermal radiation agrees with the semiclassical 
result when we choose an integration constant properly by physical reasoning. 
Furthermore, we also solve the Wheeler-De Witt equation in the region between
the inner Cauchy horizon and the outer apparent horizon, and show that the mass loss rate of 
an evaporating black hole has the same expression. 
The present study is the natural generalization of the case of a Schwarzschild  black hole 
to that of a charged Reissner-Nordstrom black hole.
\end{abstract}

\newpage
\pagestyle{plain}
\pagenumbering{arabic}


\rm
\section{Introduction}

Black holes have been thus far a quite fascinating research field from various
points of view. On the one hand, from the observational viewpoint, the black holes
are a very important constituent of the universe. It is nowdays believed that
we have super-massive black holes with masses ranging from a million to a billion
solar masses at the centers of many of galaxies including our Milky Way galaxy.
Much smaller black holes could be also formed at the beginning of the universe,
but might be evaporated away up to now \cite{Hawking}.

From the theoretical point of view, black holes provide an intriguing arena to 
explore the challenges posed by the reconciliation between general relativity
and quantum field theory. When enough energy density is concentrated in a small region 
in space, it naturally collapses to form a black hole. Thus, in the high energy regime, 
an elementary particle metamorphoses into a black hole. However, there is a sharp contradiction 
between black holes and elemetary particles. For instance, the density of states in gravity, 
which is deduced by the Bekenstein-Hawking entropy formula, is different from the density 
of states in any renormalizable quantum field theory \cite{Shomer}. To put differently, at
the Planck scale, any quantum field theory reduces to a conformal field theory in the same 
spacetime dimension whereas the Bekenstein-Hawking entropy formula for black holes implies 
that gravity does to a conformal field theory in one less spacetime dimension \cite{Aharony}. 
This observation has culminated in success of the AdS/CFT correspondence \cite{Maldacena}.

A quantum-mechanical treatment of matter in the background of a classical black hole
has indicated that the black hole radiates as a black body with a certain Hawking temperature \cite{Hawking}.
Since the black hole radiation involves a mixture of gravity and quantum mechanics, the study of
black holes seems to lead us into the territory of quatum gravity, which is the missing link 
in a complete picture of the fundamental forces in nature.

In our recent study \cite{Oda0}, based on a canonical formalism for a spherically symmetric geometry
\cite{Hajicek, Hosoya2}, we have considered the black hole radiation of a Schwarzshild black hole 
both inside and outside the apparent horizon in quantum gravity where it was shown that the mass loss rate 
due to the black hole radiation is equal to that evaluated by Hawking in the semiclassical approximation 
\cite{Tomimatsu, Hosoya2}. \footnote{See also related works \cite{Hosoya3, Oda1}.}  
Then, it is natural to ask ourselves whether our formulation 
can be extended to a charged black hole, that is, the Reissner-Nordstrom black hole, or not. In this article, 
we will present the affirmative answer to this question and show that the mass loss rate of an evaporating 
black hole with an electric charge also coincides with the result obtained in the semiclassical approach
as in a Schwarzschild black hole.

This article is organized as follows: In the next section, on the basis of the canonical formalism of a system 
with a spherically symmetric black hole, which holds in the region bounded by the outer horizon and the spatial 
infinity or in the region bounded by the space-time singularity and the inner Cauchy horizon,
we calculate the mass loss rate due to black hole radiation. In Section 3, we do the same job for the region 
bounded by the inner and apparent horizons. The final section is devoted to a conclusion.

\section{Formalism}

The four-dimensional action with which we start takes the following form: 
\footnote{We mainly follow notation and conventions by Misner et al.'s textbook \cite{MTW}, 
for instance, the flat Minkowski metric
$\eta_{\mu\nu} = diag(-, +, +, +)$, the Riemann curvature tensor $R^\mu \, _{\nu\alpha\beta} = 
\partial_\alpha \Gamma^\mu_{\nu\beta} - \partial_\beta \Gamma^\mu_{\nu\alpha} + \Gamma^\mu_{\sigma\alpha} 
\Gamma^\sigma_{\nu\beta} - \Gamma^\mu_{\sigma\beta} \Gamma^\sigma_{\nu\alpha}$, and the Ricci tensor 
$R_{\mu\nu} = R^\alpha \, _{\mu\alpha\nu}$. Throughout this article, we adopt the natural units $c = \hbar = G = 1$. 
In this units, all quantities become dimensionless.}
\begin{eqnarray}
S = \int d^4 x \sqrt{- ^{(4)}g} \left[ \frac{1}{16 \pi} \, ^{(4)}R - \frac{1}{4 \pi} \, ^{(4)}g^{\mu\nu} 
(D_\mu \Phi)^\dagger D_\nu \Phi - \frac{1}{16 \pi} F_{\mu\nu} F^{\mu\nu} \right],
\label{Action 1}
\end{eqnarray}
where $\Phi$ is a complex scalar field and its covariant derivative is given by
\begin{eqnarray}
D_\mu \Phi = \partial_\mu \Phi + i e A_\mu \Phi,
\label{Cov-derivative}
\end{eqnarray}
with $e$ and $A_\mu$ being the electric charge of $\Phi$ and the $U(1)$ gauge field, respectively.
Moreover, $F_{\mu\nu}$ is the field strength defined as
\begin{eqnarray}
F_{\mu\nu} = \partial_\mu A_\nu - \partial_\nu A_\mu.
\label{Field strength}
\end{eqnarray}
To clarify the four-dimensional meaning we put the suffix $(4)$ in front of the metric tensor and
the scalar curvature. As a final note, the Greek indices $\mu, \nu, \cdots$ take the four-dimensional
values 0, 1, 2 and 3 whereas the Latin ones $a, b, \cdots$ do the two-dimensional values 0 and 1.
Of course, it is straightforward to include the other matter fields as well as the cosmological constant
in the action (\ref{Action 1}) even if we limit ourselves to the action for simplicity.  

The most general spherically symmetric ansatz for the four-dimensional line element is of form
\begin{eqnarray}
^{(4)}ds^2 &=& ^{(4)}g_{\mu\nu} dx^\mu dx^\nu, \nonumber\\
&=& g_{ab} dx^a dx^b + \phi^2 ( d\theta^2 + \sin^2 \theta d\varphi^2 ),
\label{Line element}
\end{eqnarray}
where the two-dimensional metric $g_{ab}$ and the radial function $\phi$ are the functions of only the 
two-dimensional coordinates $x^a$. The substitution of the ansatz (\ref{Line element}) into the action
(\ref{Action 1}) and then integration over the angular variables $(\theta, \varphi)$ produces the
two-dimensional effective action
\begin{eqnarray}
S &=& \frac{1}{2} \int d^2 x \sqrt{-g} \left[ 1 + g^{ab} \partial_a \phi \partial_b \phi 
+ \frac{1}{2} R \phi^2 \right]            \nonumber\\
&-& \int d^2 x \sqrt{-g} \left[ \phi^2 g^{ab} (D_a \Phi)^\dagger D_b \Phi 
+ \frac{1}{4} \phi^2 F_{ab} F^{ab} \right],
\label{2D action}
\end{eqnarray}
where we have assumed that $A_a$ and $\Phi$ are also the functions of the two-dimensional coordinates $x^a$
and set $A_\theta = A_\varphi = 0$.

Next let us rewrite the action (\ref{2D action}) in the form of a linear combination of the first-class constraints. 
Since in this section we wish to consider the region bounded by the outer apparent horizon and the spatial infinity, 
or the region bounded by the spacetime singularity and the inner Cauchy apparent horizon, as usual 
we regard the $x^0$ coordinate as time to cover the region by spacelike hypersurfaces.
Then, the appropriate ADM splitting of (1+1)-dimensional spacetime is given by \cite{Hajicek}
\begin{eqnarray}
g_{ab} = \left(
    \begin{array}{cc}
      \frac{\beta^2}{\gamma} - \alpha^2 & \beta \\
      \beta & \gamma
    \end{array}
  \right).
\label{ADM}
\end{eqnarray}
The normal unit vector $n^a$ which is orthogonal to the hypersurfaces $x^0 = const$ reads
\begin{eqnarray}
n^a = \left( \frac{1}{\alpha}, \, - \frac{\beta}{\alpha \gamma} \right).
\label{Normal unit}
\end{eqnarray}
The trace of the extrinsic curvature and the scalar curvature are given by
\begin{eqnarray}
K &=& \frac{\dot{\gamma}}{2 \alpha \gamma} - \frac{\beta'}{\alpha \gamma}
+ \frac{\beta}{2 \alpha \gamma^2} \gamma',             \nonumber\\
R &=& 2 n^a \partial_a K + 2 K^2 - \frac{2}{\alpha \sqrt{\gamma}} \partial_1
\left( \frac{\alpha'}{\sqrt{\gamma}} \right),
\label{Extrinsic 1}
\end{eqnarray}
where $\frac{\partial}{\partial x^0} = \partial_0$ and $\frac{\partial}{\partial x^1} = \partial_1$
are also denoted as an overdot and a prime, respectively.

With the help of these equations,
one can cast the action (\ref{2D action}) to the form
\begin{eqnarray}
S &\equiv& \int d^2 x L                  \nonumber\\
&=& \int d^2 x \Biggl[ \frac{1}{2} \alpha \sqrt{\gamma} \Biggl\{ 1 - (n^a \partial_a \phi)^2
+ \frac{1}{\gamma} (\phi')^2 - K n^a \partial_a (\phi^2) 
+ \frac{\alpha'}{\alpha \gamma} \partial_1 (\phi^2) \Biggr\}     \nonumber\\
&+& \alpha \sqrt{\gamma} \phi^2 \left\{ | n^a D_a \Phi |^2 - \frac{1}{\gamma} | D_1 \Phi |^2 \right\}
+ \frac{1}{2} \alpha \sqrt{\gamma} \phi^2 E^2 \Biggr]        \nonumber\\
&+& \int d^2 x \left[ \frac{1}{2} \partial_a ( \alpha \sqrt{\gamma} K n^a \phi^2 )
- \frac{1}{2} \partial_1 \left( \frac{\alpha'}{\sqrt{\gamma}} \phi^2 \right) \right],
\label{2D action 2}
\end{eqnarray}
where we have defined $E$ as
\begin{eqnarray}
E = \frac{1}{\sqrt{-g}} F_{01} = \frac{1}{\alpha \sqrt{\gamma}} (\dot{A}_1 - A'_0).
\label{E}
\end{eqnarray}

Now the differentiation of the action (\ref{2D action 2}) with respect to the time derivative of
the canonical variables $\Phi ( \Phi^\dagger ), \phi, \gamma$ and $A_1$ leads to the corresponding 
canonical conjugate momenta $\pi_\Phi ( \pi_{\Phi^\dagger} ), \pi_\phi, \pi_\gamma$ and $\pi_A$
as follows:
\begin{eqnarray}
\pi_\Phi &=& \sqrt{\gamma} \phi^2 n^a ( D_a \Phi )^\dagger, \quad 
\pi_{\Phi^\dagger} = \sqrt{\gamma} \phi^2 n^a D_a \Phi, \quad
\pi_\phi = - \sqrt{\gamma} n^a \partial_a \phi - \sqrt{\gamma} K \phi,  \nonumber\\
\pi_\gamma &=& - \frac{1}{4 \sqrt{\gamma}}  n^a \partial_a (\phi^2), \quad
\pi_A = \phi^2 E.
\label{Momenta}
\end{eqnarray}
Then, the Hamiltonian, which is defined as
\begin{eqnarray}
H = \int d x^1 \left( \pi_\Phi \dot{\Phi} + \pi_{\Phi^\dagger} \dot{\Phi}^\dagger 
+ \pi_\phi \dot{\phi} + \pi_\gamma \dot{\gamma} + \pi_A \dot{A_1} - L \right),
\label{Hamiltonian 1}
\end{eqnarray}
is expressed in terms of a linear combination of constraints as expected from diffeomorphism
invariance
\begin{eqnarray}
H = \int d x^1 \left( \alpha H_0 + \beta H_1 + A_0 H_2 \right),
\label{Hamiltonian 2}
\end{eqnarray}
where $\alpha, \beta$ and $A_0$ are non-dynamical Lagrange multiplier fields, and the Hamiltonian 
constraint, the momentum one and the constraint associated with the $U(1)$ gauge transformation are respectively
given by
\begin{eqnarray}
H_0 &=& \frac{1}{\sqrt{\gamma} \phi^2} \pi_\Phi \pi_{\Phi^\dagger} - \frac{\sqrt{\gamma}}{2} 
- \frac{(\phi')^2}{2 \sqrt{\gamma}} + \partial_1 \left( \frac{\partial_1 (\phi^2)}{2 \sqrt{\gamma}} \right)
+ \frac{\phi^2}{\sqrt{\gamma}} | D_1 \Phi |^2                                
\nonumber\\
&-& \frac{2 \sqrt{\gamma}}{\phi} \pi_\phi \pi_\gamma
+ \frac{2 \gamma \sqrt{\gamma}}{\phi^2} \pi_\gamma^2 + \frac{\sqrt{\gamma}}{2 \phi^2} \pi_A^2,
\nonumber\\
H_1 &=& \frac{1}{\gamma} \left[ \pi_\Phi D_1 \Phi + \pi_{\Phi^\dagger} (D_1 \Phi)^\dagger \right]
+ \frac{1}{\gamma} \pi_\phi \phi' - 2 \pi'_\gamma - \frac{1}{\gamma} \pi_\gamma \gamma',
\nonumber\\
H_2 &=& - i e \left( \pi_\Phi \Phi - \pi_{\Phi^\dagger} \Phi^\dagger \right) - \pi'_A. 
\label{Constraints 1}
\end{eqnarray}

We are now in a position to apply this canonical formalism for understanding 
the Hawking radiation of a charged black hole in quantum gravity. 
We make use of the ingoing Vaidya metric \cite{Vaidya} to describe a dynamical black hole.
We therefore define the two-dimensional coordinates $x^a$ as
\begin{eqnarray}
x^a = ( x^0, x^1 ) = ( v - r, r ),
\label{Advanced time}
\end{eqnarray}
where $v$ is the advanced time coordinate. Next, let us fix the two-dimensional diffeomorphisms
by the gauge conditions
\begin{eqnarray}
g_{ab} = \left(
    \begin{array}{cc}
      \frac{\beta^2}{\gamma} - \alpha^2 & \beta \\
      \beta & \gamma
    \end{array}
  \right)
       = \left(
    \begin{array}{cc}
      - \left( 1 - \frac{2M}{r} + \frac{Q^2}{r^2} \right) & \frac{2M}{r} - \frac{Q^2}{r^2} \\
      \frac{2M}{r} - \frac{Q^2}{r^2}  & 1 + \frac{2M}{r} - \frac{Q^2}{r^2}
    \end{array}
  \right),
\label{Gauge}
\end{eqnarray}
where $M = M (x^0, x^1)$  and $Q = Q (x^0, x^1)$ play the role of the mass function and the
electric charge function of the charged black hole, respectively.
From the gauge conditions (\ref{Gauge}), the two-dimensional line element takes the form
of the Vaidya metric \cite{Vaidya}
\begin{eqnarray}
ds^2 &=& g_{ab} dx^a dx^b, \nonumber\\
&=& - \left( 1 - \frac{2M}{r} + \frac{Q^2}{r^2} \right) dv^2 + 2 dv dr.
\label{2D line element}
\end{eqnarray}

Furthermore, for the $U(1)$ gauge transformation, we shall take the gauge condition
\begin{eqnarray}
A_0 = A_1 = \frac{Q}{r},
\label{U(1) gauge}
\end{eqnarray}
which becomes 
\begin{eqnarray}
A_v = \frac{Q}{r}, \quad A_r = 0,
\label{U(1) gauge 2}
\end{eqnarray}
where we have used the definition (\ref{Advanced time}). 

For a dynamical black hole, we make use of the local definition of the horizon,
namely, the apparent horizon, rather than the global one, the event horizon.
Note that in a charged black hole there are two apparent horizons which are defined as
\begin{eqnarray}
r_\pm = M \pm \sqrt{M^2 - Q^2},
\label{Apparent horizon}
\end{eqnarray}
where $r_+$ and $r_-$ are respectively the outer and the inner horizons. In the extremal black hole $M = Q$,
the two horizons coalesce into one horizon. For simplicity, in this article we will confine ourselves to
the non-extremal black hole $M \neq Q$ since the extremal case can be treated in a similar way even if 
we will later use the property that the Hawking temperature of the extremal black hole is zero.

Since we take account of a massless scalar field which follows along the null geodesics, 
it seems to be physically reasonable to assume that the scalar field $\Phi$ depends on only the null coordinate $v$,
by which the mass function $M$ and the charge function $Q$ also become the function of the advanced 
time coordinate $v$. Furthermore, the radial function $\phi$ is assumed to be a radial coordinate $r$.
Thus, we take the following assumptions  
\begin{eqnarray}
\Phi \approx \Phi(v),  \quad M \approx M(v), \quad Q \approx Q(v),  \quad \phi \approx r.
\label{Assumption}
\end{eqnarray}
Henceforth, we shall use the simbol $\approx$ to indicate the equalities holding under the
assumptions (\ref{Assumption}). 
Under the assumption $Q \approx Q(v)$, the gauge condition (\ref{U(1) gauge}) (or (\ref{U(1) gauge 2})) 
gives rise to an electric field
\begin{eqnarray}
E \approx \frac{Q}{r^2}.
\label{EM strength}
\end{eqnarray}
It will turn out that the assumptions (\ref{Assumption}) play a critical role
in simplifying the diffeomorphism constraints and lead to a solvable model of a quantum 
black hole. 

Now we will turn our attention to quantum theory. To do that, under the assumptions (\ref{Assumption}),
using the gauge conditions (\ref{Gauge}) and (\ref{U(1) gauge 2}), let us first simplify the canonical 
conjugate momenta (\ref{Momenta}) 
\begin{eqnarray}
\pi_\Phi &\approx& \phi^2 \left( \partial_v \Phi^\dagger - i e \frac{Q}{\phi} \Phi^\dagger \right), \quad 
\pi_{\Phi^\dagger} \approx \phi^2 \left( \partial_v \Phi + i e \frac{Q}{\phi} \Phi \right), \nonumber\\ 
\pi_\phi &\approx& \frac{1}{1 + \frac{2M}{\phi} - \frac{Q^2}{\phi^2}} \left[ \partial_v M 
- \frac{Q \partial_v Q}{\phi} + \frac{2M^2}{\phi^2} + \frac{Q^2}{\phi^2} \left(1 - \frac{M}{\phi}\right) \right],  \nonumber\\
\pi_\gamma &\approx& \frac{1}{1 + \frac{2M}{\phi} - \frac{Q^2}{\phi^2}} \left( M - \frac{Q^2}{2\phi} \right), \quad
\pi_A  \approx Q.
\label{Momenta 2}
\end{eqnarray}
Then, after some calculations, it turns out that the momentum constraint becomes identical with the Hamiltonian constraint up to
an irrelevant overall factor
\begin{eqnarray}
\sqrt{\gamma} H_0 &\approx& \gamma H_1      \nonumber\\
&\approx& \frac{2}{\phi^2} \pi_\Phi \pi_{\Phi^\dagger} -  \left( 1 + \frac{2M}{\phi} - \frac{Q^2}{\phi^2} \right) \pi_\phi
+ \frac{2M^2}{\phi^2} + \frac{Q^2}{\phi^2} \left( 1 - \frac{M}{\phi} \right).
\label{Constraints 3}
\end{eqnarray}
This compatibility between the momentum and the Hamiltonian constraints justifies the assumptions (\ref{Assumption}) 
in quantum gravity. The reduction of the dynamical degree of freedom associated with the radial function $\phi$ can be understood
as a result of the momentum constraint. In addition to it, the first-class constraint coming from the $U(1)$ gauge transformation
takes the form
\begin{eqnarray}
H_2 \approx - i e \left( \pi_\Phi \Phi - \pi_{\Phi^\dagger} \Phi^\dagger \right) - \partial_v Q. 
\label{U(1)-Constraint 1}
\end{eqnarray}
 
The constraint (\ref{Constraints 3}) as an operator equation on the wave functional $\Psi$ gives rise to 
the Wheeler-De Witt equation
\begin{eqnarray}
\left[ - \frac{2}{\phi^2} \frac{\partial^2}{\partial \Phi \partial \Phi^\dagger} + 
i \left( 1 + \frac{2M}{\phi} - \frac{Q^2}{\phi^2} \right) \frac{\partial}{\partial \phi} 
+ \frac{2M^2}{\phi^2} + \frac{Q^2}{\phi^2} \left( 1 - \frac{M}{\phi} \right) \right] \Psi = 0.
\label{WDW 1}
\end{eqnarray}
Now it is easy to find a special solution of the Wheeler-De Witt equation (\ref{WDW 1}) by the method of separation
of variables. The result is given by
\begin{eqnarray}
\Psi = \left( B  e^{\sqrt{A} (\Phi + \Phi^\dagger)} + C  e^{- \sqrt{A} (\Phi + \Phi^\dagger)} \right)  
e^{ i \left[ M \log \phi + \frac{2A - M^2 - Q^2}{2 \sqrt{M^2 + Q^2}} \log \frac{\phi - \omega_-}{\phi - \omega_+} 
- \frac{M}{2} \log (\phi - \omega_+)(\phi - \omega_-) \right] }, 
\label{WDW-solution 1}
\end{eqnarray}
where $A, B$ and $C$ are integration constants. We have defined $\omega_\pm = - M \pm \sqrt{M^2+Q^2}$.

Next, one must impose the gauge constraint $H_2$ on the wave functional $\Psi$
\begin{eqnarray}
\left[ e \left( \Phi^\dagger \frac{\partial}{\partial \Phi^\dagger} - \Phi \frac{\partial}{\partial \Phi} \right) 
- \partial_v Q \right] \Psi = 0. 
\label{U(1)-WDW 1}
\end{eqnarray}
It is easy to check that when the solution (\ref{WDW-solution 1}) is substituted into this equation, 
Eq. (\ref{U(1)-WDW 1}) is satisfied only when
\begin{eqnarray}
\partial_v Q = 0, \quad \Phi^\dagger = \Phi.  
\label{U(1)-WDW-solution 1}
\end{eqnarray}
Alternatively, the second equation can be expressed as $e = 0$. In other words, the matter field must be neutral. 
We will see later that Eq. (\ref{U(1)-WDW-solution 1}) is consistent with the field equations.

Provided that the expectation value $< \cal{O} >$ of an operator
$\cal{O}$ is defined as
\begin{eqnarray}
< {\cal{O}} > = \frac{1}{\int d \Phi |\Psi|^2} \int d \Phi \Psi^* {\cal{O}} \Psi,
\label{Exp 1}
\end{eqnarray}
by using (\ref{Momenta 2}) or (\ref{Constraints 3}) one can calculate the expectation value of mass loss rate $< \partial_v M >$ 
\begin{eqnarray}
< \partial_v M > = - \frac{2 A}{\phi^2}.  
\label{Mass-loss 1}
\end{eqnarray}
In our previous article \cite{Oda0}, it has been shown that in the case of a Schwarzschild black hole ($Q = 0$), this result precisely 
coincides with that obtained by Hawking in his semiclassical approach. It will be shown later that the result (\ref{Mass-loss 1}) also
agrees with the semiclassical result of the mass loss rate of a charged black hole due to the Hawking radiation.

Here let us check whether the above assumptions (\ref{Assumption}) are compatible 
with the field equations. The field equations obtained from the action (\ref{2D action})
are given by
\begin{eqnarray}
&{}& - \frac{2}{\phi} \nabla_a \nabla_b \phi + \frac{2}{\phi} g_{ab} \nabla_c \nabla^c \phi
+ \frac{1}{\phi^2} g_{ab} \partial_c \phi \partial^c \phi -  \frac{1}{\phi^2} g_{ab}  
\nonumber\\
&=& 4 \left[ (D_a \Phi)^\dagger D_b \Phi - \frac{1}{2} g_{ab} (D_c \Phi)^\dagger D^c \Phi
+ \frac{1}{2} F_{ac} F_b \, ^c - \frac{1}{8} g_{ab} F_{cd}^2 \right],
\nonumber\\
&{}& \frac{1}{\sqrt{-g}} \partial_a \left( \sqrt{-g} g^{ab} \partial_b \phi \right)
- \frac{1}{2} R \phi = - 2 \phi (D_a \Phi)^\dagger D^a \Phi - \frac{1}{2} \phi F_{ab}^2,
\nonumber\\
&{}& D_a \left( \sqrt{-g} \phi^2 g^{ab} D_b \Phi \right) = 0,
\nonumber\\
&{}& \frac{1}{\sqrt{-g}} \partial_b \left( \sqrt{-g} \phi^2 F^{ab} \right)
= i e \phi^2 g^{ab} \left[ \Phi^\dagger D_b \Phi - \Phi (D_b \Phi)^\dagger \right].
\label{Field equations}
\end{eqnarray}
The Vaidya metric (\ref{2D line element}) gives us the metric tensor
\begin{eqnarray}
g_{ab} = \left(
    \begin{array}{cc}
      g_{vv} & g_{vr} \\
      g_{rv} & g_{rr}
    \end{array}
  \right)
       = \left(
    \begin{array}{cc}
      - \left( 1 - \frac{2M}{r} + \frac{Q^2}{r^2} \right) & 1 \\
      1 & 0
    \end{array}
  \right).
\label{Vaidya}
\end{eqnarray}
To check the compatibility of the assumptions (\ref{Assumption}) with the field
equations (\ref{Field equations}), let us make an ansatz that the variables have the form
\begin{eqnarray}
M \approx M(v), \quad Q \approx Q(v), \quad \phi \approx r,
\label{Ansatz}
\end{eqnarray}
but the scalar field is the function of both $v$ and $r$, i.e., $\Phi \approx \Phi(v, r)$.
The first field equation in (\ref{Field equations}) is satisfied if the following relations
hold:
\begin{eqnarray}
\partial_r \Phi \approx 0, \quad \Phi^\dagger \approx \Phi, \quad \partial_v Q \approx 0, 
\quad \partial_v \Phi \approx \frac{1}{r} \sqrt{\frac{\partial_v M}{2}},
\label{Solution 1}
\end{eqnarray}
where note that $ \Phi^\dagger = \Phi$ means $e = 0$ at the same time. Namely, the field
equations require that $\Phi$ is a real field and the electric charge $Q$ is constant. This fact
is consistent with Eq. (\ref{U(1)-WDW-solution 1}).

Then, the second and last equations in (\ref{Field equations}) are satisfied automatically. Finally, 
using Eq. (\ref{Solution 1}), the third field equation in (\ref{Field equations}) gives us the equation
\begin{eqnarray}
2 r \partial_v \partial_r \Phi + 2 \partial_v \Phi + r \left( 1 - \frac{2 M}{r} + \frac{Q^2}{r^2} \right) 
\partial_r^2 \Phi \approx 0.
\label{Solution 2}
\end{eqnarray}
This equation (\ref{Solution 2}) together with Eq. (\ref{Solution 1}) requires us
to limit ourselves to working with the vicinity of the apparent horizons (\ref{Apparent horizon})
\cite{Oda0}. In other words, for the consistency of the field equations, in addition to the assumptions
(\ref{Assumption}), one has to supplement one more assumption
\begin{eqnarray}
r \approx r_\pm \equiv M \pm \sqrt{M^2 - Q^2}.
\label{Assumption 2}
\end{eqnarray}

Consequently, the field equations become 
\begin{eqnarray}
\Phi^\dagger &\approx& \Phi, \quad e \approx 0, \quad Q \approx const.,   \nonumber\\
\partial_r \Phi &\approx& 0,     \nonumber\\
\partial_v \Phi &\approx& \frac{1}{r} \sqrt{\frac{\partial_v M}{2}} \approx \frac{1}{r_\pm} \sqrt{\frac{\partial_v M}{2}},  
\nonumber\\
\partial_r \partial_v \Phi &\approx& \partial_v \partial_r \Phi \approx - \frac{1}{r^2} \sqrt{\frac{\partial_v M}{2}}
\approx - \frac{1}{r_\pm^2} \sqrt{\frac{\partial_v M}{2}}.
\label{Solution 3}
\end{eqnarray}
Then, the solution is found to be
\begin{eqnarray}
\Phi (v, r) = \left( 1 - \frac{2M}{r} + \frac{Q^2}{r^2} \right)^2 \frac{r_\pm^2}{r_\pm^2 - Q^2} \frac{1}{4 \sqrt{2 \partial_v M}} 
+ \int^v dv \frac{1}{r_\pm} \sqrt{\frac{\partial_v M}{2}}.
\label{Solution 4}
\end{eqnarray}
As a result, we have
\begin{eqnarray}
\Phi(v, r) \approx \int^v dv \frac{1}{r_\pm} \sqrt{\frac{\partial_v M}{2}},
\label{Solution 5}
\end{eqnarray}
which means that one can set $\Phi(v, r) \approx \Phi(v)$ in the vicinity of the apparent horizons.
In this sense, the assumptions (\ref{Assumption}), if the assumption (\ref{Assumption 2}) is added, are at least 
classically consistent with the field equations (\ref{Field equations}) as long as the matter field is electrically neutral
and the black hole charge $Q$ is a constant.

\section{Black hole radiation in the region bounded between the inner and outer horizons}

In this section, we wish to consider the Hawking radiation in the region bounded between the inner and outer 
apparent horizons.
A peculiar characteristic feature is that in this region the Killing vector $\frac{\partial}{\partial t}$ becomes
spacelike, so that one must foliate this region with a family of spacelike hypersurfaces such as $r = const$.
We have already constructed such a canonical formalism reflecting this fact \cite{Hosoya2, Oda0} so we can apply it for the
present purpose.  The argument proceeds in a perfectly analogous way to the case treated in the previous section 
so let us explain the content in a concise manner.

In this region, the ADM splitting of (1+1)-dimensional spacetime is of form \cite{Hosoya2, Oda0}
\begin{eqnarray}
g_{ab} = \left(
    \begin{array}{cc}
      \gamma & \beta \\
      \beta & \frac{\beta^2}{\gamma} - \alpha^2
    \end{array}
  \right).
\label{ADM 2}
\end{eqnarray}
The normal unit vector $n^a$ to the Cauchy hypersurfaces $x^0 = const$ reads
\begin{eqnarray}
n^a = \left( \frac{\beta}{\alpha \gamma}, \, - \frac{1}{\alpha} \right).
\label{Normal unit 2}
\end{eqnarray}
The trace of the extrinsic curvature is calculated to be
\begin{eqnarray}
K = - \frac{\gamma'}{2 \alpha \gamma} + \frac{\dot{\beta}}{\alpha \gamma}
- \frac{\beta}{2 \alpha \gamma^2} \dot{\gamma}.
\label{Extrinsic 3}
\end{eqnarray}

Then, the canonical conjugate momenta $p_\Phi, p_{\Phi^\dagger}, p_\phi, p_\gamma$ and $p_A$ are given by
\begin{eqnarray}
p_\Phi &\approx& - \phi^2 \left( \partial_v \Phi^\dagger - i e \frac{Q}{\phi} \Phi^\dagger \right), \quad 
p_{\Phi^\dagger} \approx - \phi^2 \left( \partial_v \Phi + i e \frac{Q}{\phi} \Phi \right), \nonumber\\ 
p_\phi &\approx& \frac{1}{1 - \frac{2M}{\phi} + \frac{Q^2}{\phi^2}} \partial_v M  + 1 - \frac{M}{\phi},  \nonumber\\
p_\gamma &\approx& - \frac{\phi}{2}, \quad   p_A  \approx - Q,
\label{Momenta 3}
\end{eqnarray}
where we have used $Q = const$.  
The momentum constraint turns out to become proportional to the Hamiltonian constraint again
\footnote{Precisely speaking, since we consider the real matter field,
we should set $p_\Phi = p_{\Phi^\dagger}$, but for a generality we do not do so 
since the obtained result is in essence unchanged.}
\begin{eqnarray}
\sqrt{\gamma} H_0 &\approx& - \gamma H_1      \nonumber\\
&\approx& \frac{2}{\phi^2} p_\Phi p_{\Phi^\dagger} - \left( 1 - \frac{2M}{\phi} + \frac{Q^2}{\phi^2} \right) 
\left( p_\phi - 1 + \frac{M}{\phi} \right).
\label{Constraints 4}
\end{eqnarray}
This compatibility between the momentum and the Hamiltonian constraints justifies the assumptions (\ref{Assumption}) 
in quantum gravity as well. The gauge constraint $H_2$ becomes vanishing when the matter field is real $\Phi^\dagger 
= \Phi$ and the black hole charge $Q$ is a constant. 

An imposition of the constraint (\ref{Constraints 4}) as an operator equation on the wave functional $\Psi$
produces the Wheeler-De Witt equation 
\begin{eqnarray}
\left[ - \frac{2}{\phi^2} \frac{\partial^2}{\partial \Phi \partial \Phi^\dagger} 
- \left( 1 - \frac{2M}{\phi} + \frac{Q^2}{\phi^2} \right) 
\left( -i \frac{\partial}{\partial \phi} - 1 + \frac{M}{\phi} \right) \right] \Psi = 0.
\label{WDW 3}
\end{eqnarray}
Then, a special solution of the Wheeler-De Witt equation (\ref{WDW 3}) is given by
\begin{eqnarray}
\Psi = \left( B  e^{\sqrt{A} (\Phi + \Phi^\dagger)} + C  e^{- \sqrt{A} (\Phi + \Phi^\dagger)} \right)  
e^{ i \left[ \phi - M \log \phi - \frac{A}{\sqrt{M^2 - Q^2}} \log \frac{\phi - r_+}{\phi - r_-} \right] }, 
\label{WDW-solution 3}
\end{eqnarray}
where $A, B$ and $C$ are integration constants. 
As before, the expectation value of mass loss rate $< \partial_v M >$ is easily calculated to be
\begin{eqnarray}
< \partial_v M > = - \frac{2 A}{\phi^2},  
\label{Mass-loss 2}
\end{eqnarray}
which is the same expression as Eq. (\ref{Mass-loss 1}).
After the condition (\ref{Assumption 2}) is inserted to Eq. (\ref{Mass-loss 2}) (or Eq. (\ref{Mass-loss 1})), 
we obtain
\begin{eqnarray}
< \partial_v M > = - \frac{2 A}{r_\pm^2}
\label{Mass-loss 3}
\end{eqnarray}
This result holding in the region $r_- < r < r_+$ precisely coincides with that in the region $0 < r < r_-$ 
or $r_+ < r < \infty$.

Here there is an important caveat. It is well-known that the inner Cauchy horizon $r = r_-$ is unstable and inflates rapidly 
under the metric perturbations, what is called, "mass inflation" \cite{Poisson, Ori, Brady}. Then, our formalism 
should be applied only for the outer apparent horizon $r = r_+$ since we assume the stability of the apparent horizons implicitly.

The remaining work is to fix the integration "constant" $A$ in the sense that $A$ is independent of $\Phi$ and $\phi$.
In the present formalism, however, it is difficult to determine the vaule of $A$ in principle since the physical
state $\Psi$ is not uniquely defined in the whole spacetime region and the norm $\int d \Phi |\Psi|^2$ is not finite 
so that we will rely on the physical reasoning.

First of all, let us recall that the mass loss rate of a charged black hole should vanish for the extremal black hole 
since the extremal black hole has the vanishing Hawking temperature because of the formula $T_H \propto r_+ - r_-$ 
for the charged black hole.  
Next, notice that since $A$ is a dimensionless "constant", it must take 
the form of $(\frac{r_+ - r_-}{r_+})^n$ where $n$ is a positive number. Note that the possibility of $(\frac{r_+ - r_-}{r_-})^n$ 
is excluded since it becomes divergent in the limit of $Q \rightarrow 0$. \footnote{Incidentally, in the case of 
a Schwarzschild black hole, there is only one apparent horizon, so the integration constant $A$ must be strictly a constant.}
Finally, Stefan's law determines $n = 4$ since the energy flux for a massless field is proportional to $T_H^4$. 
Accordingly, we have
\begin{eqnarray}
A = c \left( \frac{r_+ - r_-}{r_+} \right)^4,
\label{}
\end{eqnarray}
where $c$ is some numerical constant. Hence, the expectation value of mass loss rate is given by 
\begin{eqnarray}
< \partial_v M > &=& - 2 c \, \frac{1}{\phi^2} \left( \frac{r_+ - r_-}{r_+} \right)^4   \nonumber\\
&\approx& - 2 c \, \frac{(r_+ - r_-)^4}{r_+^6},
\label{Mass-loss 4}
\end{eqnarray}
where we have used $\phi \approx r_+$ as explained above. It is remarkable that this result for a charged black hole
exactly coincides with the result obtained in the semiclassical approach up to an overall constant.

\section{Conclusion}

In this article, we have derived the mass loss rate of a charged black hole
in a purely quantum-mechanical way, but in order to fix the integration constant 
we have had to reply on the physical reasoning, for instance, the vanishing Hawking 
temperature for the extremal black hole. The difficulty of determining the integration 
constants such as $A$ is connected with the difficulty of determining the physical state 
uniquely which holds in the whole spacetime region and the norm is not well-defined. 
In spite of such a difficulty, it is surprising that the present formalism offers 
a much greater degree of flexibility in that it explains the the mass loss rate of 
not only a charged black hole but also a Schwarzschild black hole due to the Hawking radiation.

In a recent progress on black hole physics, many of important studies are closely
related to the quantum-mechanical behavior of the (apparent or event) horizon of 
a black hole. In such a situation, the present formalism seems to provide for
a useful playground. Although there remain a lot of problems to be solved,
for instance, understanding of the proper definition of the norm and the meaning 
of the physical state etc., the approach adopted in this article appears to deserve careful 
study in future.

\begin{flushleft}
{\bf Acknowledgements}
\end{flushleft}

This work is supported in part by the Grant-in-Aid for Scientific 
Research (C) No. 25400262 from the Japan Ministry of Education, Culture, 
Sports, Science and Technology.


\end{document}